\documentclass{elsart}
\usepackage{epsfig}

\begin{document}

\begin{frontmatter}

\title{Experimental study of high frequency stochastic resonance in Chua circuits}
\author{Iacyel Gomes$^{1}$, Claudio R. Mirasso$^{1}$, Ra\'{u}l Toral$^{1,2}$}  
\author{and O. Calvo$^1$}
\address{$^1$ Departament de F{\'\i}sica, Universitat de les Illes Balears, E-07122 
Palma de Mallorca, Spain }
\address{$^2$ Instituto Mediterr\'aneo de Estudios Avanzados, IMEDEA (CSIC-UIB), 
E-07122 Palma de Mallorca, Spain }

\begin{abstract}
We study the stochastic resonance phenomenon occurring in electronic Chua
circuits operating in the chaotic regime when the forcing signal is modulated
at relatively high frequency, of the order of magnitude of the main Chua frequency in the absence of forcing ($f_o$).  In all the cases, a clear maximum in the signal-to-noise ratio for an intermediate noise level is observed. When
modulating with a frequency smaller than $f_o$  we also observe a resonance at
a harmonic of the external frequency which is closer to $f_o$ or to one of its
harmonic. If the  modulating frequency is larger than $f_o$ we only observe the
resonance at the modulating frequency.
\end{abstract}

\end{frontmatter}

Stochastic resonance is a nonlinear effect that accounts for the optimum
response of a dynamical system to an external forcing at a precise value of the
noise level. Since its original proposal as a mechanism to explain the observed
periodicity of the Earth's ice ages, it has been found in a very large
number of systems of physical, as well as biological, chemical, etc,
interest\cite{BSV81,NN81,GHJM98}. A related effect is that of coherence
resonance where the optimum response (in the sense of optimal periodicity) also
occurs for a precise value of the noise level but in the absence of an external
forcing\cite{PK97}. Both effects have been observed in excitable as well as
chaotic systems \cite{PTMCG01,CMT01}, amongst others. 

In this paper we experimentally study stochastic resonance on a Chua circuit
operating in the chaotic regime. We consider the transitions from one scroll
attractor to its mirror image induced by an external periodic forcing when the
system operates under the effect of noise. At variance with similar studies, we
chose a relatively large frequency, $f$, for the external modulation, of the
order of magnitude of the main frequency, $f_o$, of the unperturbed Chua
circuit. For all frequencies considered we observe a maximum in the
signal-to-noise ratio at frequency $f$, a clear  indication that stochastic
resonance is taking place. Moreover, for forcing frequencies $f$ smaller than the unperturbed Chua frequency $f_o$ we observe stochastic resonance also at the
harmonics of the modulating frequency $f$.
\vspace*{0.75cm}

\begin{figure}[!ht]
\centerline{\makebox{\epsfig{figure=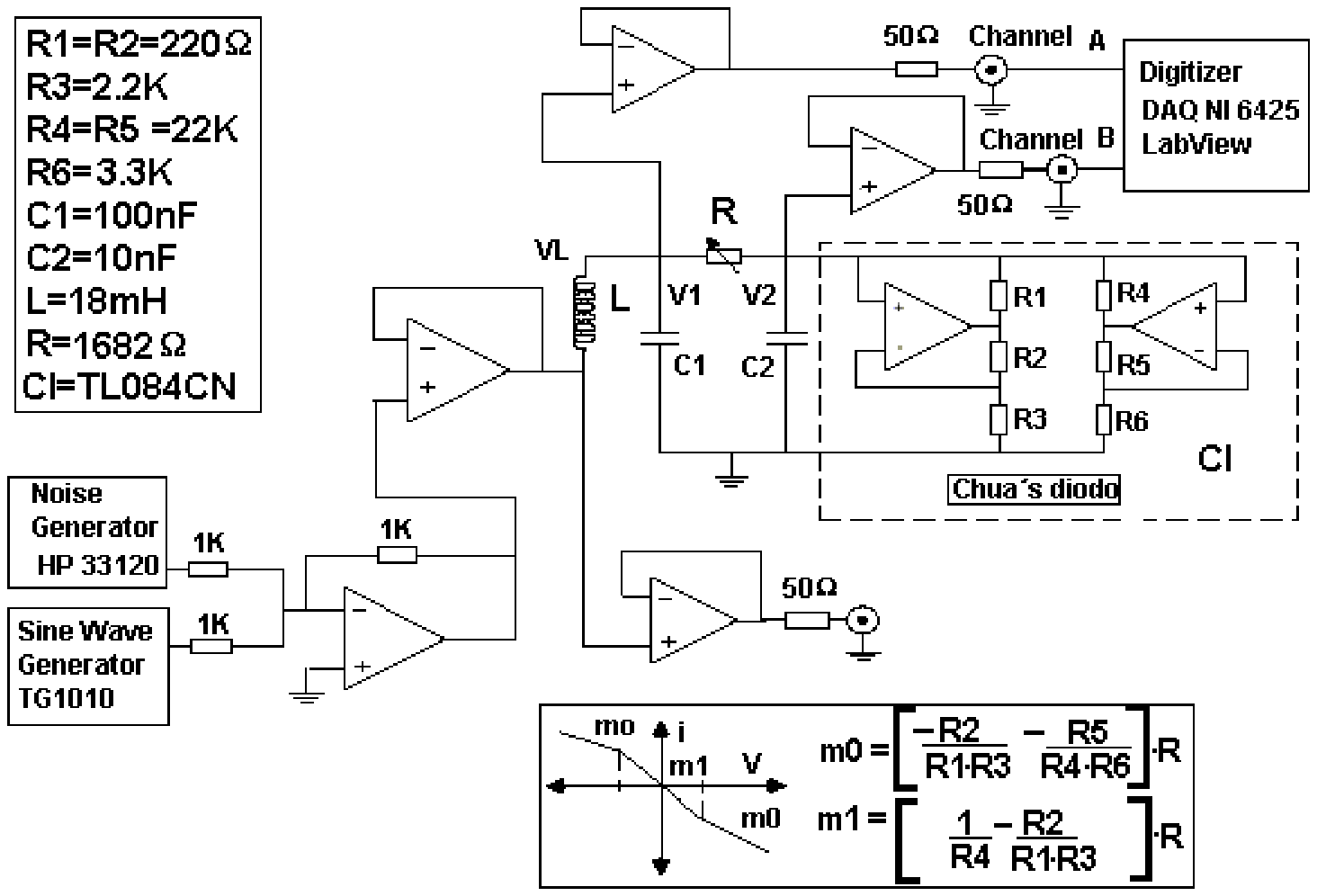,width=9cm,height=7cm,angle=0}
\epsfig{figure=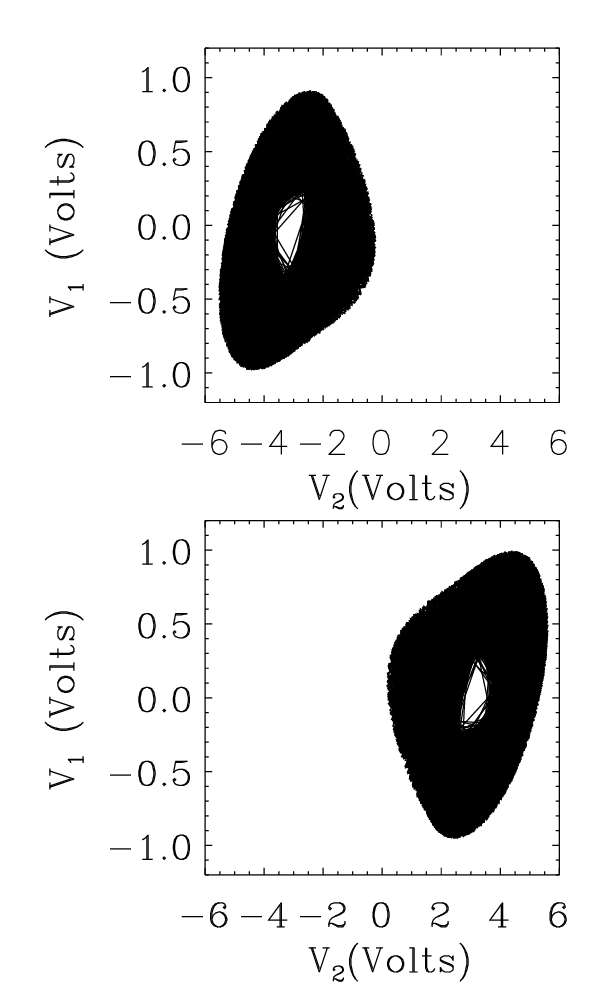,width=5cm,height=7cm,angle=0}}}
\vspace{0.5truecm}\caption{\label{fig1} Left panel: implemented Chua Circuit;
right panel: single scroll chaotic attractor (upper panel) and mirror image (lower panel).}
\end{figure}

Our Chua circuit is schematically shown in fig. 1. A digital acquisition board
from National Instruments NI-DAQ plugged into a computer was used to digitize
the signal at a sampling rate of $10$KHz. A Labview program controlled the
board in a continuous acquisition mode and an HP--33120 function generator was
used to provide the noise signals of intensity $D$ ranging from $0$ to $5$
V[rms]. A signal generator is used to force the system with periodic forcing of
the form $V(t)=V_o \sin (2 \pi f t)$, being $V_o$ the amplitude of the force and
$f$ its frequency.

This Chua circuit can be described in terms of three non-linear first order
differential equations\cite{Chua}, including those for the time evolution of
the output voltages $V_1(t) $ and $V_2(t)$ (see fig. 1). These equations
predict that, for some range of parameters, and in the absence of external
forcing, the system can have several unstable fixed points such that the pair
$(V_1(t),V_2(t))$ oscillates chaotically in time around the corresponding
dynamical attractors. For the set of parameters we have chosen in our
experiments (see inset of fig. 1), two unstable fixed points lead to an
isolated {\sl single scroll} attractor and its mirror image (see fig. 1). The
main frequency of the oscillations of the unperturbed circuit about each
attractor is $f_o \sim 2.7$ KHz. This is clearly indicated as a maximum in the
power spectrum of the system, see fig. 2. Besides the characteristic frequency
$f_o$, one can observe the appearance of several peaks at other relatively high
frequencies, not far from $f_o$, 

\begin{figure}[!ht]
\centerline{\epsfig{figure=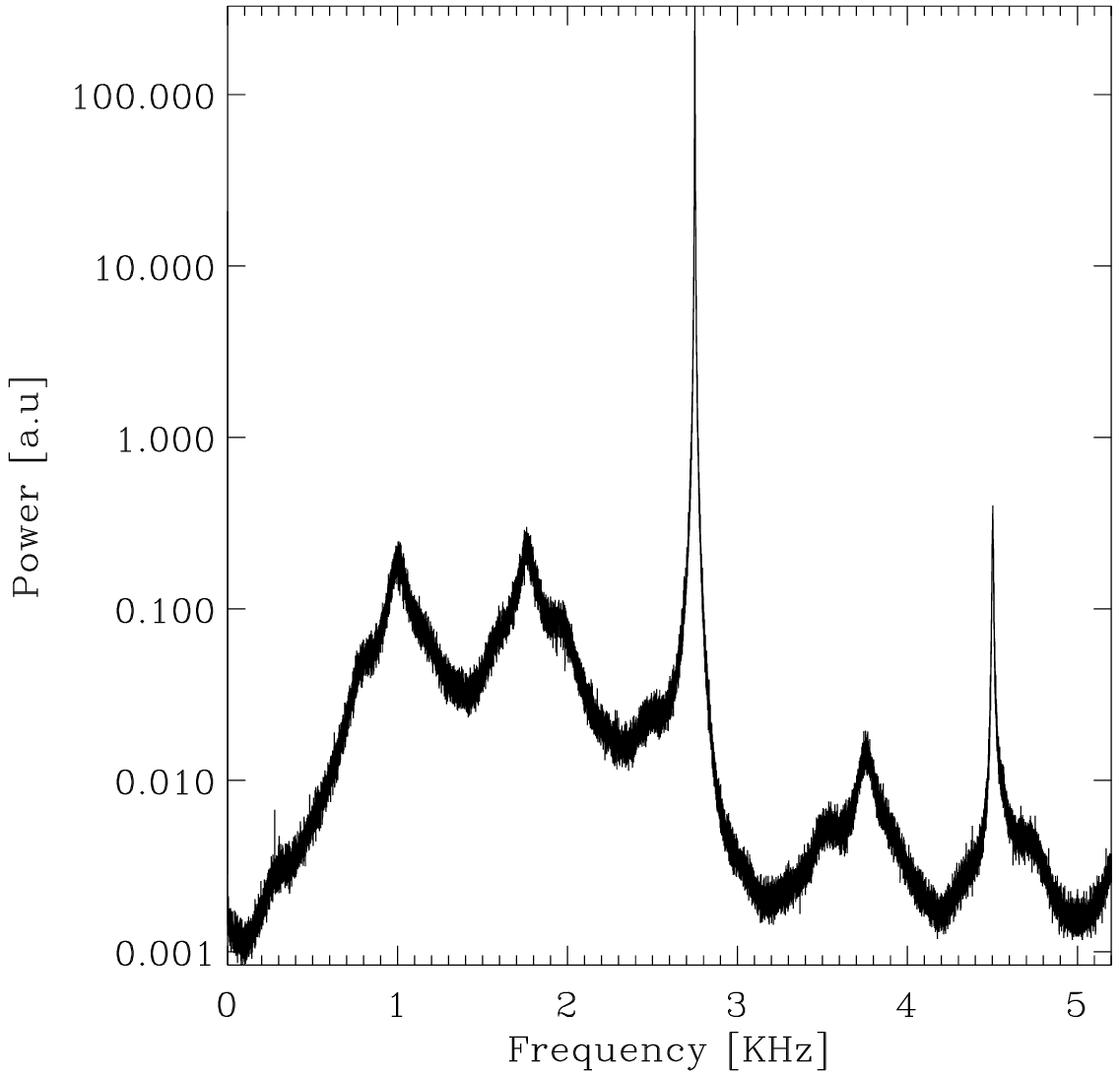,width=7cm,height=5cm,angle=0}
\epsfig{figure=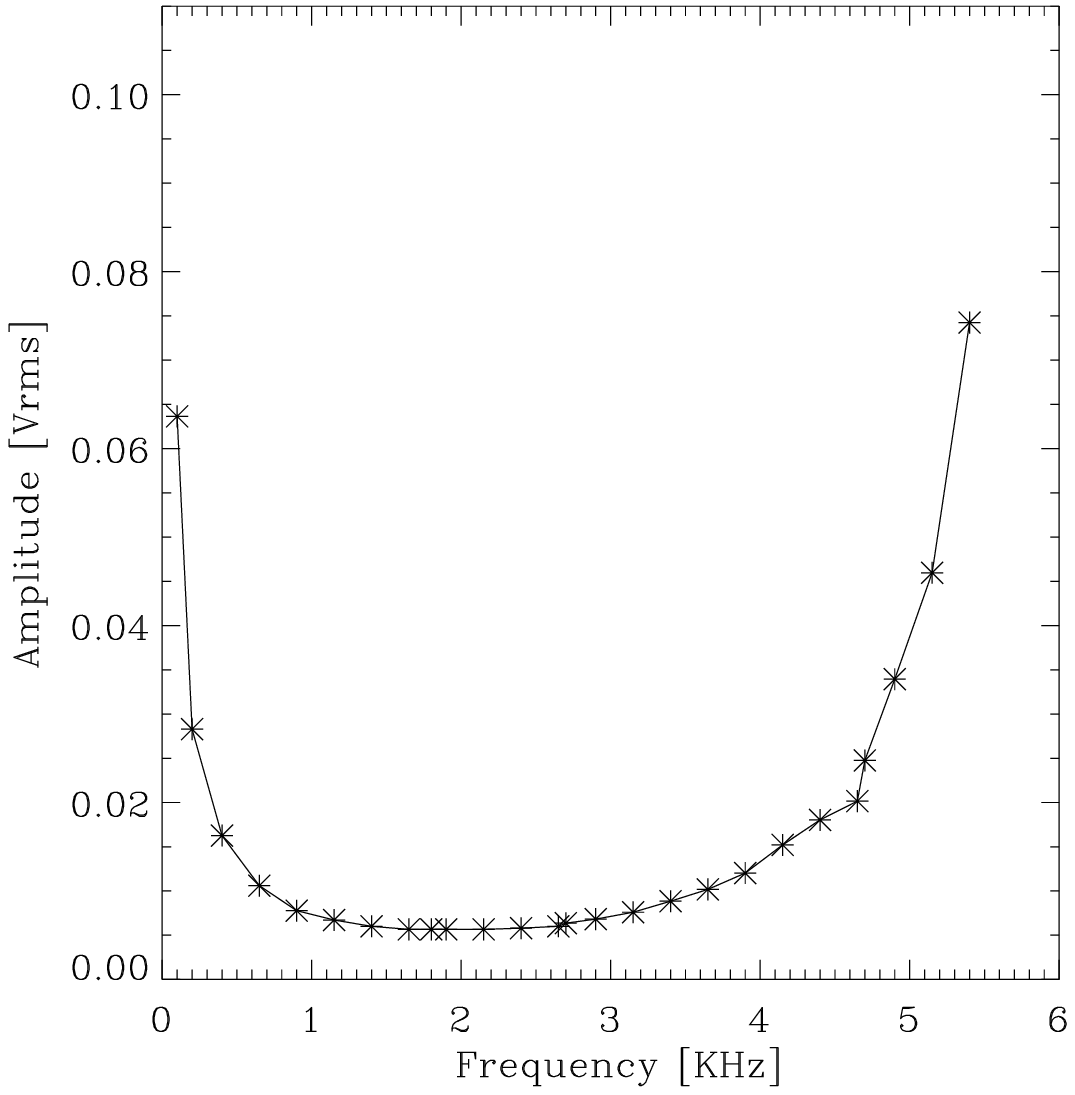,width=7cm,height=5cm,angle=0}}
\vspace{0.5truecm}\caption{\label{fig2} Left panel: Power spectrum of the system in the absence of 
the external forcing. Right panel: Threshold value for the amplitude of the 
external forcing $V_o$ for the appearance of jumps between the single scroll attractors as a function of the forcing frequency $f$.}
\end{figure}

Note that for the chosen set of parameters, and without any external forcing,
there is no possibility to jump from one attractor to the other, and which
attractor is chosen depends exclusively on initial conditions. The presence of
an external periodic forcing, however, can induce jumps between the attractors.
For such an external forcing (but still in the absence of noise) a
minimum value $V_{\rm m in}$ of its amplitude $V_o$ is required to induce those
jumps. This minimum value is a degree of the sensitivity of the system to the
external forcing and it is also plotted in fig. 2 as a function of the frequency $f$
of the forcing. It can be seen that close to the main unperturbed Chua
frequency $f_o$ the amplitude is minimum and that it increases for smaller and
larger frequencies.

With this information at hand, for each forcing frequency $f$ we now set the
amplitude of the external forcing close, but below, the minimum value $V_{\rm m
in}$ defined before. Under these conditions, the inclusion of noise of very
small amplitude is able to induce jumps between the attractors. It is in the
periodicity of these jumps where one can observe a resonance effect with
respect to the noise intensity. We have focused our study in the cases where
the external frequency coincides with some of the major peaks in the power
spectrum of the unperturbed Chua system as shown in fig. 2. Namely, we have
chosen $f=0.9$ KHz, $f=1.8$ KHz, $f=2.7$ KHz and $f=4.5$ KHz. Note that the possibility of
having stochastic resonance in a chaotic circuit with two attractors under the
presence of noise has been shown theoretically and numerically by Anishchenko
and coworkers\cite{ASC92,ANMS99}. This case is the the so--called {\sl
deterministic stochastic resonance} which appears in deterministic chaotic
systems close to the onset of intermittency, i.e., a situation different to the
one considered here.

\begin{figure}[!ht]
\centerline{\makebox{\epsfig{figure=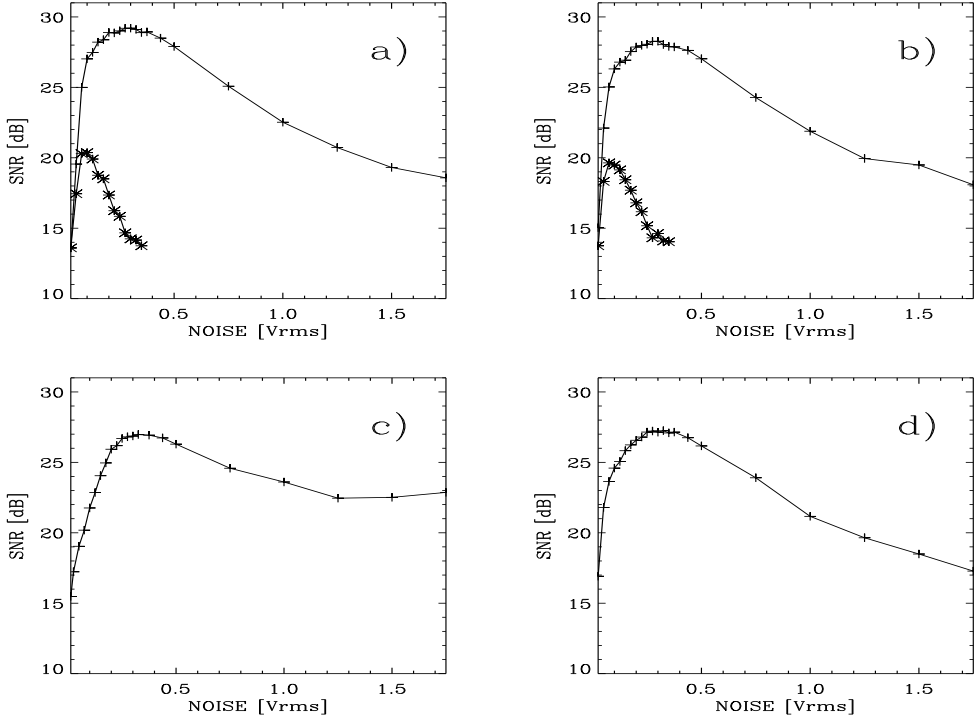,width=14cm,height=10cm,angle=0}}}
\vspace{0.5truecm}\caption{\label{fig3} Signal-to-noise ratio (SNR) vs. noise level for different forcing amplitudes: (a) $f$=
0.9 KHz, $V_o$=7 mV$_{\rm{rms}}$; (b) $f$=1.8 KHz, $V_o$=5 mV$_{\rm{rms}}$; (c) $f$=2.7 KHz,
$V_o$=6.4 mV$_{\rm{rms}}$ and d) $f$=4.7 KHz, $V_o$=25 mV$_{\rm{rms}}$.
In panels (a) and (b) crosses stand for the response at the forcing frequency ç
and stars for the response at its third harmonic. 
Solid lines are included to aid the eye.}
\end{figure}

Under modulation and noise, the power spectrum varies the relative height of
their peaks. We characterize the response by the usual signal to noise ratio
indicator. In fig. 3 we plot the results we obtain when modulating with
different frequencies. In all the panels it can be clearly seen that even at
these high frequencies a maximum of the signal-to-noise ratio is observed for
an intermediate noise level, the most important signature of stochastic
resonance. Moreover, the maxima occur for approximately the same noise level
independently of the forcing frequency. Interestingly, when looking at the
response at frequencies $f$ corresponding to other peaks in the power spectrum,
we observe that for frequencies smaller than the main Chua frequency $f_o$ the
system also exhibits stochastic resonance at harmonics of this forcing
frequency $f$. It appears that stochastic resonance appears only for those
harmonics whose frequency is closer to $f_o$ or to a harmonic of $f_o$. For
example, in fig.4 panel a) another stochastic resonance effect is observed in
the third harmonic of the forcing frequency (in this case $3f=2.7$ KHz $\approx
f_o$) while in panel b) an stochastic resonance effect occurs at $3f=5.4$ KHz,
which corresponds to $2 f_o$. On the contrary, when the external frequency is
larger than $f_o$, we only observe stochastic resonance at the forcing
frequency.

In conclusion, we have experimentally shown that stochastic resonance in an
electronic Chua circuit occurs even when the system is modulated at relatively
high frequencies, as compared to the usual stochastic resonance in which the
forcing frequency is small compared with any internal frequency. When forcing
with a frequency smaller than the main unperturbed Chua  frequency we have also
shown that the resonance occurs not only at the forcing frequency $f$ but also
at harmonics of $f$ which are close to the main unpertubed Chua frequency $f_o$
or to a harmonic of it.

The work is supported by MCyT (Spain) and FEDER, projects BFM2001-0341-C02-01,
BMF2000-1108 and BFM2002-04369. I.G. acknowledges support form the Agencia
Espa\~{n}ola de Cooperaci\'{o}n Internacional.

\end{document}